# The Trusted Server – a secure computational environment for privacy compliant evaluations on plain personal data.

Nikolaus von Bomhard 1*¶, Bernd Ahlborn 1, Catherine Mason 1, Ulrich Mansmann 1

1 Department of Medical Information Sciences, Biometrics, and Epidemiology

Ludwig-Maximilian's University, Munich, Germany

* Corresponding author

bomhard@ibe.med.uni-muenchen.de (nvb)

## Abstract:

A growing framework of legal and ethical requirements limit scientific and commercial evaluation of personal data. Typically, pseudonymization, encryption, or methods of distributed computing try to protect individual privacy. However, computational infrastructures still depend on human system administrators. This introduces severe security risks and has strong impact on privacy: system administrators have unlimited access to the computers that they manage including encryption keys and pseudonymization-tables. Distributed computing and data obfuscation technologies reduce but do not eliminate the risk of privacy leakage by administrators. They produce higher implementation effort and possible data quality degradation. This paper proposes the *Trusted Server* as an alternative approach that provides a sealed and inaccessible computational environment in a cryptographically strict sense. During operation or by direct physical access to storage media, data stored and processed inside the *Trusted Server* can by no means be read, manipulated or leaked, other than by brute-force. Thus, secure and privacy-compliant data processing or evaluation of plain person-related data becomes possible even from multiple sources, which want their data kept mutually secret.

## 1. Introduction:

### 1.1 Background

Both scientific and commercial statistical evaluation of data in the fields of epidemiology, pharmacology, education or economics use *person-related* data containing highly sensitive private information. This comprises person-identifying data (also called person-related data like name, address, date of birth etc.), which privacy protection rules do address, as well as *person-relatable* information, which allow identifying a person by using re-identification techniques [1]. Legislation [2] and ethical conventions [3] impose strict privacy protection rules not only regarding person-related but also person-relatable information. While data evaluation may be permitted by law or consent for a certain purpose [4], it has to be ensured that any other usage of pri-vacy-related data is effectively prevented. Other areas with growing interest in privacy protection are social networks [5] or highly security relevant networks e.g. for military use [6].

## 1.2 The problem

Privacy protection in computational environments requests to protect data and computational processes from unauthorized human access. Current computational environments allow access control, data-storage and -transport protection by user-authentication and user-rights management, as well as disk- and transport-encryption. Additionally, pseudonymization permits evaluation of privacy-protected data that are readable for humans. However, none of those methods provides protection against access, infringing pseudonymization, manipulation and theft by an administrator with root-rights on involved servers. The core problem of privacy protection and data security is the need for a system administrator with unlimited rights to manage computers.

## 1.3 Existing solutions

Current solutions to this security and privacy core risk make use of data processing diversification over multiple computational instances and obfuscation techniques:

### 1.3.1 Double Coding Pseudonymization

A data source provides pseudonymized data, e.g. patients' clinical data with the identifying values replaced by pseudonyms. A trusted third party exchanges the $1^{st}$ level pseudonyms with new $2^{nd}$ level pseudonyms and forwards the medical data with the $2^{nd}$ level pseudonyms to the evaluating institution. The matching between $1^{st}$ and $2^{nd}$ level pseudonyms is kept secret at a trusted third party so no direct depseudonymization can be done by members of the data source and evaluating institutions neither accidentally or willingly [7].

### 1.3.2 Differential Privacy

Adding non-destructive randomness to real data as well as random data that look like real data obfuscates datasets. Ideally, this process – optionally combined with pseudonymization - hin-ders or eliminates the identification of the person behind these data but does not affect the statistical evaluations on certain variables [8,9].

### 1.3.3 **Secure Multiparty Computation**

This method uses encrypted data exchange and complex multi-stage algorithms allowing multiple parties to commonly evaluate a function over their respective private data without giving the other parties access to these private data. [10]

### 1.3.4 **DataShield**

Instead of aggregating data in one place where evaluations are performed, the underlying calculations are being sent to the data owners for in-place evaluation. Only results are returned and aggregated for further processing so no confidential private data ever leave the data owner's infrastructure. [11]

## 1.4 Common disadvantages

a) All methods described in section 1.3 protect data more or less against access from system administrators but share the weakness of increased effort for planning, implementation, infrastructure, administration and operation. Their complexity outgrows, as more parties will get involved.

b) Any kind of data-alteration by pseudonymization or obfuscation affects data quality. The degree of possible data degradation can be approximately quantified by applying these methods to publically available data and compare them to a direct naïve evaluation.

c) Without obfuscation there is the risk of privacy leakage even from pseudonymized data with *person-relatable* information.

## 1.5 A different approach

The human factor creates disadvantages related to the methods described in section 1.3. Therefore, a generic, widely adaptable computational environment that works without any human system intervention or possible access to internal data provides the needed solution. We call such an environment the *Trusted Server* (TS) and define its requirements for a practical implementation as follows:

### 1.5.1 Standards compliance

Hard- and software-components are commonly available and do not require low level customization or modification out of the ordinary.

### 1.5.2 Familiar operation

Setup, operation and usage is similar and comparably complex to administrating a conventional server with the same configuration.

### 1.5.3 Full transparency

The solution is fully transparent and does not work with secrets or obfuscation.

### 1.5.4 Unlimited verifiability

Users can review all components and the fully working system in any depth desired.

### 1.5.5 System inaccessibility

There is no system access neither during runtime nor after production.

### 1.5.6 Secure communication

The TS allows controlled submission of data and commands as well as controlled response.

### 1.5.7 Persistent encryption

The TS uses irrevocably encrypted storage which protects against external access by anyone at any time.

### 1.5.8 System verification

It is possible to verify the production system state is unaltered.

### 1.5.9 Backup strategy

It is possible to backup and restore a basic TS installation in a comfortable way.

### 1.6 Possible advantages of a Trusted Server

a) Data stored and processed inside TS do not need additional data- or privacy protection. Data securely uploaded to TS after sealing, does not need to be pseudonymized, obfuscated or encrypted.

b) This provides the unique possibility to store and evaluate unaltered plain person-related data even from different and mutually non-trusting sources in one single computational stage.

c) Working on plain unaltered data grants the highest information quality possible excluding any data degradation and impact on results deriving from obfuscation or pseudonymization.

d) There is no technical and administrative overhead caused by involving multiple parties, pseudonymization and obfuscation.

### 1.7 Implications

Any data uploaded to the TS after sealing by design are inevitably lost if the TS needs a new setup and have to be uploaded again. Depending on the data-amount this may cause serious delay requiring alternative concepts for securely delivering large data.

### 1.8 A working solution

Running sample applications of real world scenarios are provided on a reference implementation of the proposed TS. The TS is not just a new concept but an available stable production platform for previously impossible privacy protected data evaluation on plain unaltered personal data.

## 2. Materials and methods

### 2.1 Meeting the requirements

a) The use of exclusively freely available hardware and Free Open Source Software (FOSS) grants standards compliance, familiarity, transparency and verifiability. Our first implementation uses Debian GNU/Linux as operating system in a default installation with Linux Unified Key Setup (LUKS) and Logical Volume Manager (LVM) disk encryption. Other unixoid FOSS operating systems may qualify as well.

b) Simple shell scripts running at startup realize system inaccessibility. They remove all system user accounts, block root login, remove ssh completely, and set firewall and hosts access control to block all but https network traffic.

c) Secure communication is possible over secure and encrypted https with optional system independent user authentication.

d) Persistent encryption is the core method. Based on LUKS disk encryption a two stage sealing mechanism is established.

e) Any party concerned prior to sealing can inspect disk images of the readily prepared TS system. Further verification of the TS features follows from inspecting comprehensive logs and checksums after sealing. They prove the server's unaltered state.

f) The system disk images allow restoring the system in a fast and convenient way.

## 2.2 LUKS based system sealing and verification

During initial operating system setup, LUKS (together with LVM) enables disk encryption. LVM is secondary to understand disk encryption and the sealing process. Therefore, we omit a thorough discussion of its role. During the installation of a new Linux system with full encryption, the system disk splits into two data partitions: partition1 one for the static boot files and partition2 for the encrypted operating system, as well as other software, and user data. In fact, there is an additional 'partition' respectively logical volume for memory-swapping as well as possible additional volumes for user data or whatever. Since those logical volumes are located within the encrypted partition2 we simply discuss the boot and encrypted partition in the following:

a) After dividing the disk into two partitions, the LUKS header is written to partition2. The LUKS header consists of 8 key-slots. Each of them can store a copy of the master-key which is encrypted with a keyphrase. The keyphrase may be manually entered or automatically read from a keyfile [12]. We store the keyfile within the unencrypted boot-partition1. The master-key is used to encrypt the data area of partition2, but itself is never persistently stored anywhere (see Fig 1).

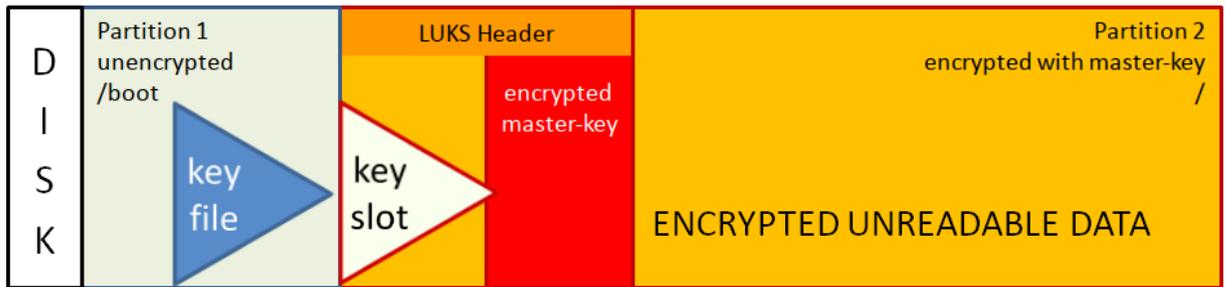

Fig 1 Initial layout of the LUKS encrypted disk

b) During the boot process, the `initrd` with the core operating system contents is loaded from the unencrypted partition1 and control moves to LUKS. Usually, a user submits now a keyphrase. Instead, the TS system reads the key-file from the unencrypted partition1 and compares it with the matching keyslot-entry in the LUKS header. With the verified passphrase it decrypts the encrypted master-key and stores it in Random Access Memory (RAM). Since data in RAM are volatile on power loss, one has to redo the decryption procedure during every system boot (see Fig 2).

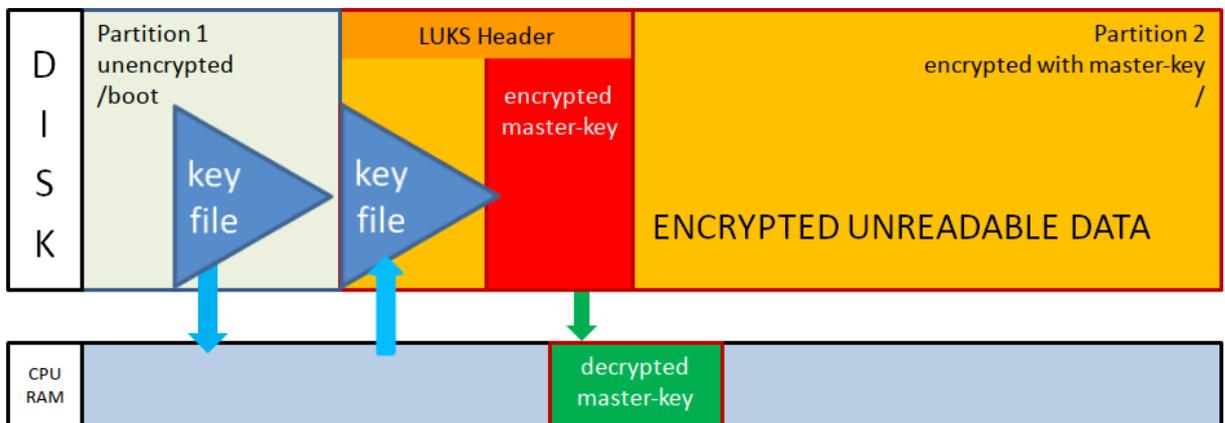

Fig 2 The master-key is decrypted using the passphrase and stored in volatile memory

c) The processor transparently reads and writes from and to partition2 using the master-key as long as the master-key resides in RAM. Data on partition2 will always be encrypted; decrypted data only exist in volatile memory (see Fig 3).

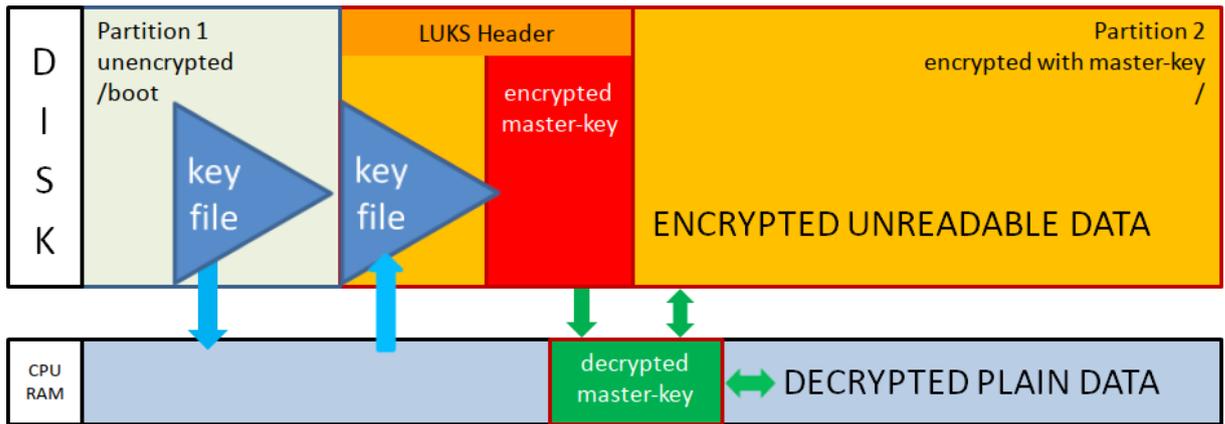

**Fig 3 Operational system state with transparent data de- and encryption**

d) The sealing process starts immediately upon booting a production ready TS and erases the LUKS keyslot as well as the encrypted master-key. The master-key still resides in volatile memory and the system remains operative but the keyfile containing the keyphrase is meaningless since neither a keyslot nor an encrypted version of the master-key exists (see Fig 4).

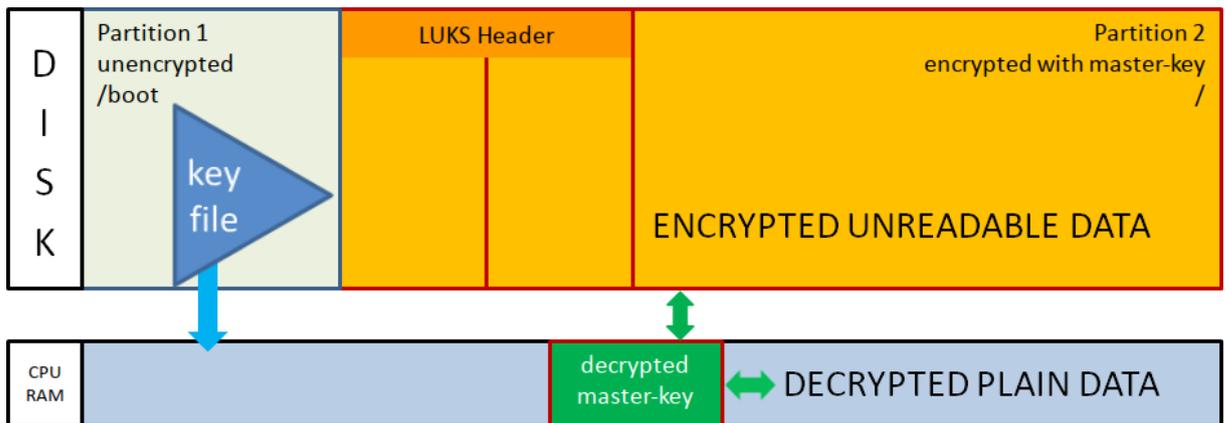

**Fig 4 Sealed operational state**

e) The master key vanishes from volatile memory If the system is rebooted or power is down (either willingly or e.g. upon theft of the server or disk). The key file still exists on the unencrypted partition1 but without the corresponding LUKS keyslot containing the encrypted master-key. The only way to decrypt partition2 is by brute force (see Fig 5).

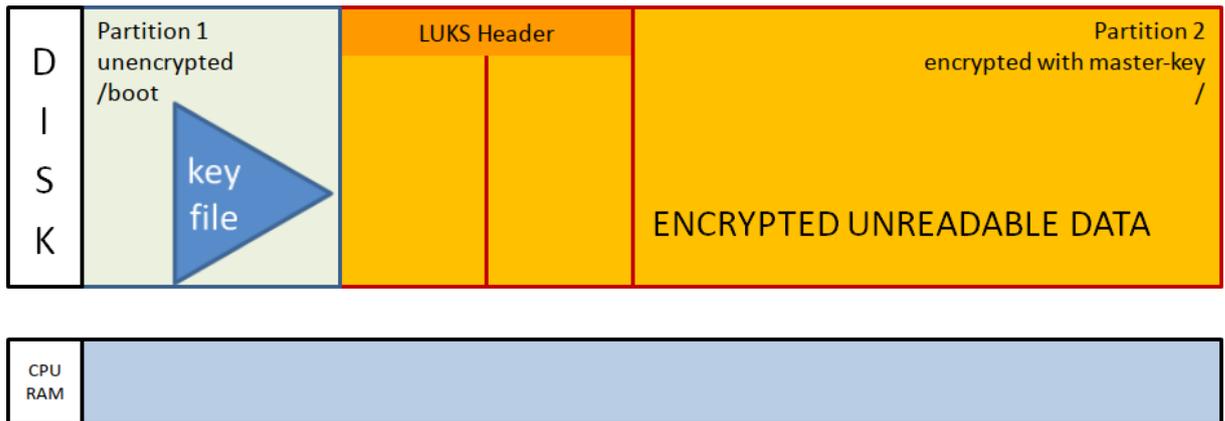

**Fig 5 Inaccessible disk state after reboot or power down**

f) As described up to now, the sealing process prevents effectively any access to the system and storage by third parties. But, it does not yet solve the basic problem. An administrator might have a backup of the LUKS header and restore it to regain disk access. The following trick overcomes this problem: We establish a two stage setup consisting of a physical server, a virtual machine hosted on it, and two LUKS-encrypted physical disks.

g) The physical server boots from disk1 and performs the sealing. After sealing, it reencrypts the second disk using the keyfile stored in that disk's partition1. LUKS reencryption creates a new master-key that is stored encrypted with the given keyfile. While the system administrator knows that keyfile he does not know the newly generated master-key. It cannot be revealed from the already sealed physical host server either.

h) Finally the physical server starts the virtual machine which boots from disk2 and performs the self-sealing process (see Fig 6) too.

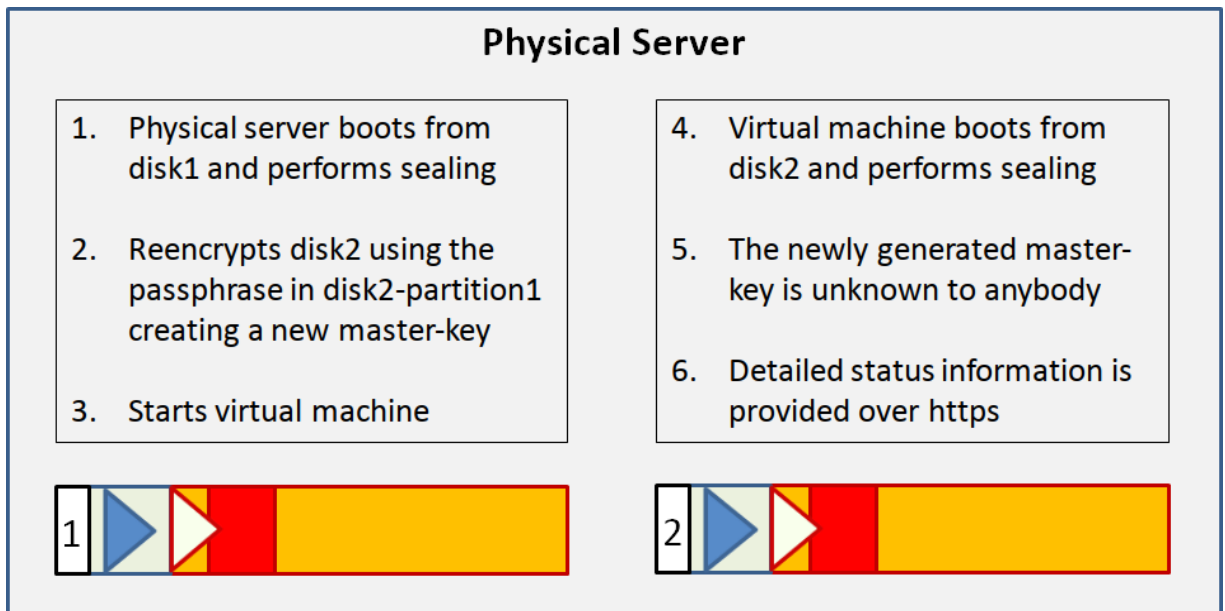

**Fig 6 The complete Trusted Server with dual stage sealing**

## 2.3 Applications and customization

a) The system administrator implements an apache2 web server configured for https traffic that provisions the sealing logs and system verification data. If required apache2 also enables secure data input and output as well as system independent user authentication.

b) Depending on the TS's further configuration and initialization procedures, ssh access is configured and secured.

c) The TS virtual machine also provides additional services and operative user applications that are needed.

d) Following good practice for configuring a server, IPtables firewall and host access control reduce access and allowed network traffic to the required minimums.

## 2.4 System verification

a) When the TS is installed, full disk images of the physical host and the virtual machine are stored in a safe place.

b) Anyone can fully inspect these disk images to validate the TS installation and state.

c) Each step of the sealing process is logged. The log-files provide comprehensive status information on the host and the virtual machine disk of the sealed TS:

- The system writes SHA-512 hashes from all files on the disks and to the sealing log.

- It lists essential configuration files in the sealing log.
- It archives configuration folders in compressed format.

d) The system publishes sealing log, system logs, and the compressed configuration archives to the (optionally access restricted) Trusted Server's website. Thus, anyone can compare the sealed state with the content of the previously disk images disclosed for verification.

## 2.5 Backup and restore

The disk images created from the host and virtual machine disks can also be used for fast restore of the Trusted Server's pre-sealing state in case of a configuration change or system maintenance.

## 2.6 Initializing production state

Simple bash-scripts perform the sealing process (section 2.2) automatically on a fully installed and purposely configured Trusted Server.

## 2.7 Initialization scripts reference

The following batch-scripts specify our Trusted Server implementation. They can be easily modified and customized. Their linear stepwise structure intends to provide easy readability of the sealing log.

### 2.7.1 Initialization scripts executed on TS-Host

init_trusted_mode.sh (manually executed by administrator)

```
set –x
## INIT TRUSTED MODE TS-HOST
## ACTIVATE SEALING AFTER REBOOT AND WRITE OUTPUT TO LOGFILE
echo '/root/init_trusted_mode_reboot.sh > /root/0_init_trusted_mode_host.log 2>&1'
>> /root/cron-reboot.sh
## -- REMOVE ONLY LOGON USER --
userdel -f trust
##REMOVE ALLOWED HOST  ACCESS PERMISSION AND VERIFY
rm /etc/hosts.allow
cat /etc/hosts.allow
```

```
cat /etc/hosts.deny
reboot
```

cron-reboot.sh (automatically triggered from /etc/crontab: @reboot)

```
#!/bin/bash
## cron-reboot TS-Host
mount /dev/sdb1 /mnt
iptables-restore /root/iptables.v4
ip6tables-restore /root/iptables.v6
## INIT TRUSTED MODE AND CREATE HOST SEALING:
/root/init_trusted_mode_reboot.sh > /root/0_init_trusted_mode_host.log 2>&1
```

init_trusted_mode._reboot.sh (called from cron-reboot.sh)

```
set –x
## INIT TRUSTED MODE TS-HOST
reboot
## SWITCH IPTABLES OUTGOING POLICY TO DROP AND DELETE SSH PERMISSION
iptables -P OUTPUT DROP
## [set line number accordingly:]
iptables -D INPUT 4
## LIST IPTABLES
iptables-save
iptables -L –n
ip6tables-save
ip6tables -L –n
cat /etc/hosts.allow
cat /etc/hosts.deny
## REMOVE SSH SERVER
apt-get -y purge openssh-server
apt-get -y autoremove
systemctl status sshd
## REMOVE ONLY LOGON USER - CREATES ERROR IF ALREADY CORRECTLY REMOVED
userdel -f trust
cat /etc/passwd
```

```
cat /etc/shadow
## REMOVE DISK ENCRYPTION KEY --
cryptsetup luksErase /dev/sda2
cryptsetup luksDump /dev/sda2
## PRINT OLD VM KEY INFORMATION
cryptsetup luksDump /dev/sdb2
## AND REENCRYPT TS-VM DISK
cryptsetup-reencrypt -v -d /mnt/keyfile -l 512 /dev/sdb2
## PRINT NEW VM DISK KEY INFORMATION
cryptsetup luksDump /dev/sdb2
## CREATE ARCHIVES OF ETC AND ROOT FOR PUBLISHING
zip -r /mnt/etc-host.zip  /etc
zip -r /mnt/root-host.zip  /root
## LIST FILES AND SHA3 CHECKSUMS
ls -RlA /
rhash -r --sha3-512 /boot
rhash -r --sha3-512 /etc
rhash -r --sha3-512 /home
rhash -r --sha3-512 /lib
rhash -r --sha3-512 /lib64
rhash -r --sha3-512 /lost+found
rhash -r --sha3-512 /media
rhash -r --sha3-512 /mnt
rhash -r --sha3-512 /opt
rhash -r --sha3-512 /root
rhash -r --sha3-512 /sbin
rhash -r --sha3-512 /srv
rhash -r --sha3-512 /tmp
## since /usr/bin has X11 -> . recursive link:
rhash    --sha3-512 /usr/bin/*
rhash -r --sha3-512 /usr/games
rhash -r --sha3-512 /usr/include
rhash -r --sha3-512 /usr/lib
rhash -r --sha3-512 /usr/local
```

```
rhash -r --sha3-512 /usr/sbin

rhash -r --sha3-512 /usr/share

rhash -r --sha3-512 /usr/src

rhash -r --sha3-512 /var

## COPY LOG TO TS-VM BOOT PARTITION

cp /root/0_init_trusted_mode_host.log /mnt

## UNMOUNT TS-VM BOOT PARTITION

umount /mnt

## START TS-VM

virsh start debian9
```

### 2.7.2  Initialization script executed on TS-VM

```
cron-reboot.sh (automatically triggered from /etc/crontab: @reboot)

    #!/bin/bash

    # cron-reboot TS-VM

    iptables-restore  /home/trust/iptables.v4

    ip6tables-restore /home/trust/iptables.v6

    ## INIT TRUSTED MODE AND CREATE VM SEALING LOG

    /home/trust/init_trusted_mode.sh > /var/www/log/1_init_trusted_mode.log 2>&1

init_trusted_mode.sh (called from cron-reboot.sh)

    set -x

    ## REMOVE HOST ACCESS PERMISSION AND VERIFY

    rm /etc/hosts.allow

    ## SWITCH IPTABLES OUTGOING POLICY TO DROP AND DELETE SSH PERMISSION

    iptables -P OUTPUT DROP

    ## [set line number accordingly:]

    iptables -D INPUT 5

    ## LIST IPTABLES AND HOST ACCESS

    iptables-save

    iptables -L –n

    ip6tables-save

    ip6tables -L –n

    cat /etc/hosts.allow
```

```
cat /etc/hosts.deny
## REMOVE SSH SERVER
apt-get -y purge openssh-server
apt-get -y autoremove
systemctl status sshd
## REMOVE ONLY LOGON USER
userdel -f trust
cat /etc/passwd
cat /etc/group
cat /etc/shadow
## REMOVE DISK ENCRYPTION KEY
cryptsetup luksErase /dev/vda2
cryptsetup luksDump /dev/vda2
## MOVE HOST LOG AND ZIP TO WEBROOT
mv /boot/0_init_trusted_mode_host.log /var/www/log
mv /boot/etc-host.zip /var/www/log
mv /boot/root-host.zip /var/www/log
chown www-data:www-data /var/www
## CREATE ARCHIVES OF /ETC AND /HOME/TRUST FOR PUBLISHING
zip -r /var/www/log/etc-vm.zip   /etc
zip -r /var/www/log/trust-vm.zip /home/trust
## CREATE LDAP LOG
date >> /var/www/log/ldap.txt && slapcat -n 0 >> /var/www/log/ldap.txt && slapcat -n 1 >> /var/www/log/ldap.txt
## SET PERMISSIONS TO APACHE2
chown -R www-data:www-data /var/www
## LIST FILES AND SHA3 CHECKSUMS
ls -RlA /
rhash -r --sha3-512 /boot
rhash -r --sha3-512 /etc
rhash -r --sha3-512 /home
rhash -r --sha3-512 /lib
rhash -r --sha3-512 /lib64
rhash -r --sha3-512 /lost+found
```

```
rhash -r --sha3-512 /media

rhash -r --sha3-512 /mnt

rhash -r --sha3-512 /opt

rhash -r --sha3-512 /root

rhash -r --sha3-512 /sbin

rhash -r --sha3-512 /srv

rhash -r --sha3-512 /tmp

## since /usr/bin has X11 -> . recursive link:

rhash    --sha3-512 /usr/bin/*

rhash -r --sha3-512 /usr/games

rhash -r --sha3-512 /usr/include

rhash -r --sha3-512 /usr/lib

rhash -r --sha3-512 /usr/local

rhash -r --sha3-512 /usr/sbin

rhash -r --sha3-512 /usr/share

rhash -r --sha3-512 /usr/src

rhash -r --sha3-512 /var

## ENABLE APACHE WEBSERVER

systemctl start apache2

## SEND MAIL

echo $(date) >> /home/trust/date.txt

mail -s "trusted server running@138.245.80.17" bomhard@ibe.med.uni-muenchen.de < home/trust/date.txt
```

## 3. Results and Discussion

### 3.1 Comparisons of the proposed methods

a) Table 1summarizes qualitative differences between the Trusted Server's generic approach and other common and well-established strategies to privacy-protected personal data evaluations We focus on server- and implementation-related but task-independent criteria: Administrative Skills, Overhead, Complexity, Adaptability, and Data Quality. It shows the Trusted Server's superiority regarding ease of implementation and usage, flexibility and negative impact on results.

**Table 1. Trusted Server versus established methods**

|  | Trusted Server | Double Coding Pseudonymization | Differential Privacy | Secure Multiparty Computation | DataShield |
|---|---|---|---|---|---|
| Administrative Skills | moderate: any average system administrator is able to follow the instructions | medium: specialized knowledge about pseudonymization software is required | high: nondestructive data obfuscation requires special skills and good planning | very high: deep knowledge in cryptography and mathematics is necessary | medium: specialized knowledge about DataShield software and setup is required |
| Overhead | very low: one sufficiently performant server for data provisioning and evaluation is all needed even by multiple parties | medium: data provisioning and evaluation must be separated in independent infrastructures plus a third party is required | moderate: the data provider must obfuscate data and evaluation has to be separated in an independent infrastructure | very high: all participants have to implement a complex and highly resource consuming computation infrastructure | high: all participants have to implement a complete software and hardware infrastructure |
| Complexity | very low: standard GNU/Linux operating system and tools and some simple shell scripts is all needed | moderate: pseudonymization software is integrated in an otherwise conventional processing chain | high: data obfuscation algorithms have to be customized for every type of use case | very high: the data processing chain has to be designed and tailored for every distinct use case | medium: distributed data processing requires careful data normalization and customized aggregation |
| Adaptability | very high: almost any technology and solution available for GNU/Linux can be used with low to zero customization | high: since pseudonymization does not affect data structures required process customization is moderate | medium: possibility and quality of data obfuscation depends on data types and evaluation purposes | very low: implementing the processing and encryption chain is singular for every use case | high: evaluation are performed on normalized but otherwise original data with standard R programs |
| Data Quality | maximum: exclusive usage of plain and unaltered data grants zero influence on results | high: in most cases pseudonymization will not, but might affect evaluation results | medium: obfuscation reduces data quality, but that may be irrelevant to evaluations | very high: since data are encrypted, but unaltered, zero degradation can be achieved | high: normalization and aggregation after processing likely will not, but could affect results |

b) Table 2 provides scenario-independent quantitative information on the additional effort for data and privacy protection caused by a Trusted Server. Comparison is made to a conventional server operating without any data protection based on typical server-lifecycle parameters (Basic installation, Customization, Initialization and Sealing, Backup and Restore, System updates) and practical usability (System stability, Performance degradation, and Resource consumption).

**Table 2. Additional effort for data and privacy protection using a Trusted Server**

| Issue | Comment | Additional effort |
|---|---|---|
| Basic installation | Two servers, host and virtual machine, have to be installed, LUKS-disk encryption needs to be set up and sealing scripts have to be installed. | about factor 3 |
| Customization | Task-specific software installation and configuration is required on the virtual machine only and in a conventional fashion. | none |
| Initialization and Sealing | Depends on installation size, disk- and system performance. Values relate to a fully functional standard Debian GNU/Linux system on two different hardware platforms. | 25 minutes on older 2CPU/8GB/SATA Laptop<br>15 minutes on 12CPU/32GB/SAS Server |
| Backup & Restore | Duration depends on disk and interface performance and installation size. Any data uploaded after sealing at least decryption keys have to be uploaded again after sealing. | + second disk restore<br>+ sealing<br>+ data or key upload |
| System update | Full restore and sealing is needed, update times itself are equal to unsecured server but have to be applied to host and virtual machine. | + restore<br>+ double updates<br>+ sealing |
| System stability | No instability or otherwise different behavior compared to our conventional servers was observed during one year of operation on several servers. | none |
| Performance degradation | Possible impact on performance by LUKS disk encryption or the virtual machine is not observable on any modern hardware. | not observable |
| Resource consumption | Moderately better equipment is required. | + second disk<br>+ 4 GB RAM for host |

c) The Trusted Server provides a new state-of-the-art regarding security and protection. Table 3 gives an overview on typical operation-related security threats like leakage of foreign data or security corruption and general threats like theft, hacking and data transfer. The most relevant (but only slightly elevated) risk for the TS relates to data transfer.

**Table 3. Systematic risks for privacy and data security in different methods**

| Threat | Trusted Server | Double Coding Pseudonymization | Differential Privacy | Secure Multiparty Computation | DataShield |
|---|---|---|---|---|---|
| Leakage of foreign data by personnel | not possible and easily verifiable | possible if the trusted third party and evaluation party work together or if the trusted third party has access to personal data | not relevant since nobody has access to foreign plain data | depending on implementation very unlikely if possible at all, but difficult to verify | not relevant since nobody has access to foreign plain data |
| Security corruption by personnel | very difficult since the sealed and frozen system state report is disclosed for in depth verification | possible at the trusted third party | possible by leakage or manipulation of obfuscation algorithms | depending on implementation very unlikely, but difficult to verify | possible at all data providers' servers |
| Theft of disk or server | full encrypted disk without LUKS header can only be decrypted by brute force attack against the master key | if disks are full encrypted disk they can only be decrypted by brute force attack against the master key or passphrase | | | |
| Hacking | slightly higher protection than a properly secured GNU/Linux server (no user logon) | the single servers can be protected on state-of-the-art level, but every additional computation and communication stage and especially added software is a potential security risk and may introduce new vulnerabilities | | | |
| Man-In-The-Middle-Attacks on data transfer | slightly higher risk since plain personal data could be accessible | slightly lower risk since no plain but still person relatable data are transferred | lower risk since transferred data are hardly person relatable | low risk since only encrypted data is transferred | low risk since only analysis commands and non-disclosing summaries are transferred |

## 3.2 Implementation scenarios

### 3.2.1 **Privacy-protected User authentication**

Basic user authentication can be implemented using apache2's file based user- and password database. After TS sealing, no change to those files is possible except by permitting security-weakening file upload.

*Lightweight Directory Access Protocol* (LDAP) [13] replication offers a more transparent and flexible directory service for storing and authenticating user credentials. Apache2 can authenticate against any LDAP server instead of using its own user and password database. The free and open source OpenLDAP [14] reference implementation permits uni- or bidirectional synchronization of the LDAP database. The Trusted Server and one or more external primary servers work with OpenLDAP. This allows secure credential updates to a sealed Trusted Server. External non-trusted primary OpenLDAP server(s) store all user credentials. The Trusted Server's OpenLDAP instance triggers unidirectional LDAP replication from the external primary OpenLDAP server(s). The system initially and regularly during operation publishes full LDAP database-dumps on the Trusted Server website. This ensures full control that OpenLDAP contains only credible users.

Thus a Trusted Server can be used with changing access permissions to the provided services without need for a new setup and sealing. While it is possible to control, that only entitled users can access a Trusted Server's web-based service, there is no control if a certain user really accesses and uses the web-service allowing for access-restricted yet anonymous online services.

### 3.2.2 Large data storage

Only after the sealing process, person-related plain data must be uploaded to the trusted server's storage. As consequence, every change or system crash requests a new data upload. To avoid long processing times for large datasets (for example when analyzing full human genomes), encrypted disks attached to the Trusted Server before sealing carry the sensitive data. After sealing, the data provider uploads the decryption key to the Trusted Server. The disk can be newly mounted in a short time.

### 3.2.3 Intentional emergency 'backdoor'

Specific scenarios request maximum data and privacy protection as well as an opportunity for secure controlled system access. Sending an encrypted copy of the master-key created during virtual machine disk reencryption to a trusted instance allows for secure controlled system access. Splitting the encrypted master-key into several parts enhances security and control when it's parts are sent to different third parties. Only the active cooperation of all parties allows system decryption.

#### 3.2.4 Automated restore

Many professional servers provide watchdog background programs. They monitor the proper operation of the server automatically. Thus, server malfunction or unresponsiveness trigger a forced cold-reset on hardware level. The server reboots and, if configured for boot over network on disk-boot failure, automatically restores the disk images and starts the initialization scripts.

### 3.3 Usage examples

#### 3.3.1 Privacy protected Domain Name Server

Server providing Domain Name Services (DNS) store and provide matching internet domain names and corresponding *internet protocol* (IP) network addresses. Whenever a user submits an internet domain name to the internet browser, a request is sent to a DNS server to provide the IP address of the corresponding server. The DNS server gets and may store the requesting users IP address and requested domain, which can be privacy sensitive information. A Trusted Server set up as an intermediate so-called DNS proxy server redirects requests to a public DNS server, providing its own network address together with the requested domain name and forwarding the returned network address to the original requesting client. Person-related clients' IP addresses are not submitted to the public DNS server.

#### 3.3.2 Yao's millionaires' problem

In 1982 Andrew C. Yao introduced the Millionaires' Problem to theoretical informatics: "Two millionaires wish to know who is richer; however, they do not want to find out inadvertently any additional information about each other's wealth. How can they carry out such a conversation?" [15]. Yao's solution relies on complex multiparty algorithms and is one of the initial formulations of secure multiparty computation. The Trusted Server permits implementing an extremely simple solution: It uploads data over a SSL-encrypted web form containing two fields, one for the name and one for the value of assets along with a submission button. On every input, the TS adds the name-value pair to a table in human readable form, perfectly protected by its privacy design. A script sorts the table by value and publishes only the names to a text file on the Trusted Server's website.

Thus, the TS not only transforms one of the challenges of theoretical informatics to common-level information technology but also provides a highly generic solution. The approach also works for large numbers of submissions without significant increase in resource consumption. It can be used either open-to-the-public or, using LDAP replication, for a closed access-controlled user group.

### 3.3.3 Anonymous webmail server

A simple transport encrypted web application with a text submission form runs on the Trusted Server. The text submission may be open to the public, or OpenLDAP authentication controls access. A nickname, comment and optionally a return email-address may be provided. Upon submission the content of the form is sent to a preconfigured email address. This can be used to provide a secure portal e.g. for whistleblowers or anonymous patients' reports in clinical studies. Combined with LDAP authentication input may be restricted to a limited user group, while retaining full anonymity at least, if the submission form is used from a public non person-related computer e.g. in an internet cafe.

## 3.4 Use case: A standard problem in epidemiological research

The example simulates the following situation:

Data collected in three centers provide the input to a prognostic model. There is a high interest in the model but reluctance to share the data openly. The data may contain sensitive information on patient mix, treatment strategies, and respective outcomes. The TS provides an elegant solution to this problem.

Utilizing R-package plumber [16] with a problem-specific R-script allows to restrict the user to the predefined R-function calls when performing the analysis and providing the results. That assures non-disclosure of information, that should not be shared openly.

For demonstration purposes and reproducibility we take the openly available dataset GBSG from the R-package mfp [17]. The dataset consists of 686 patients and we split it into three consecutive parts of about 228 patients representing the data of three different clinics. The analysis studies the influence of age (age) and the expression of progesterone receptor (PRM).

The TS provides the results of the analysis in a list which consists of the regression coefficients c1 for the fractional polynomials of age (f1) and c2 for the fractional polynomials of prm (f2) as well as the modified cumulative baseline hazards function (CBH). Both information allow to calculate group specific survival curves:  $S(t|age,prm) = \exp\{-CBH(t)*\exp[c1 \cdot f1(age)+c2 \cdot f2(prm)]\}$.

The standard CBH is a step function with jumps at each event time. Publishing the CBH in this form may allow to reidentify individual patients by observed event times. Therefore we use a smoothed form of the CBH which blurs observed event times. This deidentifying step is given in the code line www<-lowess(haz,f=0.1). This is a very practical approach that needs more thinking in a real scenario.

In the following we two R-scripts. The first R-script (plumber.R) starts the plumber server, which is remotely accessed over the apache2 proxy.

```
plumber.R
    library(plumber)
    r <- plumb("<…..demo.R>")
```

```
    r$run(port=8000)
```

The second script (demo.R) contains the analysis which mainly rely on three functions. The function getPacman attaches the library which manages the specific library attachments needed for the analysis. The function readDat concatenates the individual csv data files in the working directory to a common data object in R. The line with the hash mark before the evalrfc function is a decorator which can be interpreted by plumber defining the call the server should respond to. The function evalrfc provides the specific analysis data steps, returning the data that are responded, when the interface is called. After defining the functions, the script performs the following steps: attaching pacman, attaching the specific libraries over pacman and reading the data.

The evaluation is started and results are provided by calling the URL:

"https://<ipAddressOrDomainName>:<port>/evalrfc"

demo.R

```
    getPacman <- function()
        {
            if (!"pacman" %in% installed.packages())
            install.packages("pacman")
            library(pacman)
        }
    readDat <- function(dir)
        {
            setwd(dir)
            all_files <- list.files()
            dats <- lapply(all_files, read.csv)
            dat <- do.call(rbind, dats)
            return(dat)
        }
    #* @get /evalrfc
    evalrfc <- function()
        {
```

```
        result <- mfp(Surv(rfst, cens) ~ fp(age, df = 2, select = 0.05) + fp(prm, df
   = 4, select = 0.05), family = cox, data = dat)
        coef <- summary(result)$coefficient
        haz <- basehaz(result)
        www<-lowess(haz,f=0.1)
        res <- list(coef=coef,basehaz=www)
        return(res)
    }
   getPacman()
   p_load(mfp)
dat <- readDat("path to data")
```

## 3.5 Security considerations

### 3.5.1 **Decryption resistance**

Any grade of privacy protection and security is relative. This of course is also valid for the TS. Its grade of protection depends on the quality and irrevocability of the Virtual Machine disk2 encryption. LUKS is cryptographically strong [18] and without the key-slot keys it is impossible to decrypt the disk except by brute force - that is finding the decryption key by trial and error [19]. Successful brute-force attacks against strong encryption are limited to a few intelligence agencies in the world, if possible at all. This in most scenarios is meaningless, since those agencies will have access to the protected data anyway.

### 3.5.2 **Technical limitations**

a) Server BIOS and the CPU-Microcode are closed source and potentially contain undocumented functions and backdoors. This implies that today's real-world computing hardware cannot achieve absolute trust-to-the-last.

b) The cryptographic strength of encryption techniques for Solid-State-Disks (SSD) is currently under discussion [20]. Exploiting proprietary wear leveling technology to obtain and restore a LUKS header with deleted passphrases under rare circumstances might be possible for specialists. Therefore, SSD must not be used in a Trusted Server if maximum protection even from high-ly skilled attackers is mission-critical. Using SSD with additional hardware encryption may solve the problem. This approach still needs validation.

### 3.5.3 Tamper-resistance

After testing and approval, the system administrators activate the initialization scripts. At that point the administrators could change binaries or add scripts in the physical host or Virtual Machine. This intervention could break security, for example by sending out the secret key and LUKS-header created during Virtual Machine disk reencryption.

Thorough review of the published logs and comparison of the TS' state after installation and its state after sealing are crucial. The SHA-3 hashes and log files published on the TS's website allow to detect changes and to reveal most manipulations.

For maximum trust, transparency and control, disk images should be crated immediately before the sealing is initialized and securely provided to the concerned parties. Ideally representatives of all parties personally attend the sealing and receive their disk images. Video self-surveillance of the TS and sealing process may be disclosed over the TS website, too.

### 3.5.4 Vulnerabilities

Aside from added security by sealing, a Trusted Server shares all vulnerabilities and contact surfaces with a conventional server having an identical setup. Therefore, we recommend additional security measures:

a) Remove Gnome Virtual File System and any other auto-mounters for external storage to prevent code injection from scripts running automatically when an external USB storage or CD/DVD is inserted and external ports are needed for some reason.

b) Specific scenarios recommend to use means like hardening, creating custom kernels, to use SELinux or AppArmour. Applications installed on the Trusted Server need a careful internal security check, too.

c) Disclosing a full disk image for review allows corrupting the SSL transport encryption by a man-in-the-middle attack [21], since the private SSL key is disclosed. SSL encryption itself is not affected, as the session encryption keys are created independently from the identifying SSL key. However the identity of the Trusted Server needs approval by additional means.

d) A Cold-Boot [22], DMA [23] or removable media attack on the Trusted Server is possible either. Therefore, securing the server physically is a prerequisite e.g. by gluing or soldering in RAM-modules and physically removing or destroying CD disk drives and external ports like USB. These measures are the same as needed just to secure a conventional server with disk encryption in a given setting.

e) Additional protection and security is achieved by using a server-vault or strongroom with strict access management.

f) A physical self-destruction mechanism triggered by any human access to the server-vault may protect the TS even against the strongest attackers.

## 4. Conclusion

The TS overcomes human-centric paradigms in privacy protection concepts. All current approaches base on either trust or mistrust in single or multiple real persons. Accordingly, they establish either a network of trust, which spreads information over multiple semi-trusted instances of human-driven institutions or use complex computation schemes of fully encrypted data so nobody needs to trust anyone but himself or herself. Compared to standard non-privacy-protected solutions both approaches require highly customized workflows.

The TS may request moderately prolonged downtimes for maintenance and changes. Compared to multi stage approaches this compensates by quick and easy setup as well as minimized workflow customization.

The TS provides a conventional computational environment that grants Privacy by Design independently from any individual. Since the TS behaves - despite self-sealing and irrevocable encryption - like any standard GNU/Linux based system, it is possible to run well-established computational solutions with the highest degree of privacy.